\documentclass[aps,pra,superscriptaddress,twocolumn,showpacs,longbibliography,nofootinbib]{revtex4-1}
\usepackage{times,mathrsfs, amsmath, amsfonts, graphics, graphicx, cancel, color, theorem, bbm, mathtools, amssymb}

\usepackage[english]{babel}
\usepackage[margin=0.75in]{geometry}
\usepackage[T1]{fontenc}
\usepackage{xfrac,relsize,float}
\usepackage[unicode=true, breaklinks=false, pdfborder={0 0 1}, backref=false, colorlinks=true, linkcolor=blue, citecolor=blue, urlcolor=blue]{hyperref}
\usepackage{physics}

\definecolor{Blue}{rgb}{0,0,1}
\definecolor{Red}{rgb}{1,0,0}
\definecolor{Green}{rgb}{0,1,0}
\definecolor{darkgreen}{rgb}{0,.7,0}
\definecolor{Purp}{rgb}{.2,0,.2}
\definecolor{white}{rgb}{1,1,1}

\newcommand{\Id}{\mathbbm{1}}

\usepackage[most]{tcolorbox}

\begin{document}
\title{ A discrete memory-kernel for multi-time correlations in non-Markovian quantum processes }
\author{Mathias R. J{\o}rgensen}
    \email{matrj@fysik.dtu.dk}
    \affiliation{Department of Physics, Technical University of Denmark, 2800 Kgs. Lyngby, Denmark}
\author{and Felix A. Pollock}
    \email{felix.pollock@monash.edu}
    \affiliation{School of Physics and Astronomy, Monash University, Clayton, Victoria 3800, Australia}
\date{\today}

\begin{abstract}
Efficient simulations of the dynamics of open systems is of wide importance for quantum science and technology.
Here, we introduce a generalization of the \textit{transfer-tensor}, or discrete-time memory kernel, formalism to multi-time measurement scenarios. 
The transfer-tensor method sets out to compute the state of an open few-body quantum system at long times,
given that only short-time system trajectories are available. 
Here, we show that the transfer-tensor method can be extended to processes which include multiple interrogations (e.g. measurements) of the open system dynamics as it evolves,
allowing us to propagate high order short-time correlation functions to later times,
without further recourse to the underlying system-environment evolution.
Our approach exploits the process-tensor description of open quantum processes to represent and propagate the dynamics in terms of an object from which any multi-time correlation can be extracted.
As an illustration of the utility of the method,
we study the build-up of system-environment correlations in the paradigmatic spin-boson model,
and compute steady-state emission spectra, taking fully into account system-environment correlations present in the steady state.
\end{abstract}
\maketitle

\section{Introduction}
Modelling the dynamical properties of open quantum systems is an outstanding challenge in modern physics.
Applications range from studying impurity dynamics in ultracold quantum gases or strongly correlated materials~\cite{PhysRevLett.122.205301,PhysRevLett.123.183001,PhysRevLett.124.076601},
to modelling emission properties of organic molecules or artificial atoms~\cite{PhysRevLett.124.153602, PhysRevLett.123.167403}. While a plethora of techniques exist for the simulation of the time-dependence of observables at a single point in time~\cite{BreuerPetruccione2002,deVega2017RMP}, usually encoded in the time-dependent state of an open system, there are relatively few that are aimed at computing multi-time correlation functions in full generality. These correlation functions are critical to modelling spectral responses, such as those measured in ultrafast spectroscopy experiments~\cite{MukamelARPC1990,YuenZhou2014}, used to directly infer quantum dynamical behaviour in, for example, photosynthetic processes~\cite{FlemingNature2005}.

If the coupling between the system under study and its surrounding environment is sufficiently weak,
and the reorganization timescale of the environment sufficiently short, that memory effects can be neglected,
then multi-time correlations can be expressed in terms of two-time correlations through the so-called \textit{quantum regression theorem}~\cite{GardinerZoller2004}.
Under an additional stationarity assumption, these two-time correlations are themselves expressible in terms of single-time observable evolution~\cite{BreuerPetruccione2002}. However, for more general \textit{non-Markovian} dynamics, environment-mediated memory effects preclude such simplifications, and  correlations across multiple times are not fully determined by those across fewer~\cite{GuarnieriPRA2014,McCutcheon2016PRA,LiLi2018PhR}.

Accurately computing these correlations typically requires sophisticated numerical methods, such as those based on  path integrals~\cite{CosacchiPRB2018,Jorgensen2019PRL} and quantum Monte Carlo~\cite{ChenJCP2017-1,ChenJCP2017-2}, and other methods that rely on simplifying details of the model in question~\cite{AlonsoPRA2007}.
In many cases, these methods suffer from poor convergence with system size, with system-environment interaction strength,
and with the total simulation time.
This limits the ability to apply them to important processes involving large, complex systems or where long-time dynamics is important. 

For (single-time) state evolution, these kinds of scaling difficulties can, in some cases, be mitigated with the aid of a \textit{memory kernel}~\cite{BreuerPetruccione2002,Cohen2011PRB,VacchiniPRL2016}, a mapping that quantifies how a system's past trajectory contributes to its future evolution. Once this is determined, the system's state can be propagated in time with an efficiency that does not depend on the details of the underlying system-environment dynamics~\cite{ShiGevaJCP2003}. This is perhaps most clearly evident in the discrete-time memory kernel approach, known as the transfer-tensor method (TTM)~\cite{Cerrillo2014PRL,Pollock2018tomographically} (closely related to an earlier approach specific to the spin-boson model~\cite{GolosovJCP1999}).
The TTM can be treated as a black-box,
taking short-time non-Markovian dynamical maps, computed using some other technique, as input, and self-consistently propagating them to arbitrarily long times. 
It has been used to great success in simulating properties of systems in contact with a bosonic bath~\cite{Rosenbach_NJP2016, Buser2018PRA};
however, it is not directly applicable to the computation of multi-time correlations.

In this paper, we develop a generalization of the transfer-tensor formalism to a scenario involving sequential measurements of an non-Markovian open system as it evolves, resulting in an efficient discrete-time memory kernel method for the propagation of multi-time correlation functions. In contrast to previous multi-time memory kernel methods, for which correlations between each set of operators require a separate kernel~\cite{Ivanov2015PRA}, our approach utilizes the powerful process-tensor formalism to encode all correlations of a given order into a single positive operator~\cite{Pollock2018PRA,Jorgensen2019PRL}.
As an illustration of the utility of our generalized method,
we study the build-up of system-environment correlations in the paradigmatic spin-boson model,
and compute steady-state emission spectra, taking into account system-environment correlations present in the steady state.

\section{Background}
In this section we introduce the basic notions of open quantum dynamics and multi-time correlation functions.
Often, one is interested in the probability distribution describing observed outcomes in a sequential measurement scenario,
and we discuss how such a probability distribution can be conveniently expressed in terms of a process tensor~\cite{Pollock2018PRA}.
Lastly we outline the standard transfer-tensor method, as it applies to the propagation of the system's state.

\subsection{Open dynamics and correlation functions}
We are here concerned with the simulation of a stochastic process undergone by an open quantum system over the set of discretized times $\{t_{n} = n\delta t: n\in \mathbb{N}\}$,
where $\delta t$ defines the temporal resolution of the simulation
\footnote{For the purposes of the transfer-tensor method we derive in later sections, this resolution need not have any relation to the intrinsic timescales of the process to be simulated.
However, many techniques for computing the initial dynamics rely on $\delta t$ being sufficiently small~\cite{deVega2017RMP}.}.
In a general quantum stochastic process of the sort we consider here,
an open system ($S$) evolves along with its environment ($E$), while simultaneously being monitored and manipulated by a putative experimenter.
Its dynamics therefore has two contributions:
the first is the dynamical evolution the open system would undergo between consecutive discrete times if it were not externally perturbed,
and the second is the  influence of measurements and other external interventions.
Here, we imagine the latter are implemented sufficiently fast that they can be considered instantaneous with respect to $\delta t$;
continuous manipulations can be approximated in a controlled fashion by reducing $\delta t$ appropriately~\cite{Jorgensen2019PRL,Milz2020kolmogorovextension}.

The free dynamical evolution of an open system can always be modelled by considering that the joint $SE$ dynamics
can be described by a time evolution superoperator $\Lambda_{t, s}$ governed by the Liouville-von Neumann equation,
\begin{gather}
\partial_t\Lambda_{t, s} = \mathcal{L}_{t} \Lambda_{t, s} . 
\end{gather}
The time evolution is generated by the Liouvillian superoperator defined by $\mathcal{L}_{t}\left( \bullet \right) \equiv -i \left[H(t),\bullet \right]$,
where the Hamiltonian $H$ describes the system, the environment and their interaction.
With the initial condition $\Lambda_{s,s} = \mathcal{I}$ (the identity superoperator),
this has the formal solution $\Lambda_{t,s} = T_{\leftarrow}e^{\int_{s}^t dr\,\mathcal{L}_r}$, with $T_{\leftarrow}$ indicating time ordering of the following exponential.
We will henceforth specialise to the case where the $SE$ Liouvillian is time-independent: $\mathcal{L}_t=\mathcal{L}$ $\forall t$,
such that, with respect to our discrete time grid, we can write $\Lambda_{n} := \Lambda_{t_n, 0} = e^{n\delta t \mathcal{L}}$. In this case, we have that 
\begin{equation}
    \Pi_{n} = e^{n \delta t \mathcal{L}} \left( \Pi_{0} \right) ,
\end{equation}
where $\Pi_{n}$ denotes the full system-environment state at time $t_{n}$,
and the above equation relates this state to the full initial state $\Pi_{0}$.
Throughout we focus on the scenario in which the dynamics is generated by a time-independent Hamiltonian,
however the results we present can straightforwardly be generalized to a periodically time-dependent Hamiltonian following the approach of Pollock et al.~\cite{Pollock2018tomographically}.
Furthermore we work in units in which $\hbar=1$, such that time is measured in units of inverse energy.

It is often the case, that we want to compute operator expectation values or multi-time correlation functions.
Suppose, that the system has an observable represented by the Hermitian operator $A$ with spectral resolution $A=\sum_{k}a_{k} \ket{a_{k}}\!\bra{a_{k}}$,
then the expectation value of the operator at time $t_{n}$ is given in the Heisenberg picture by
\begin{equation}
\begin{aligned}
    \left\langle A(t_{n}) \right\rangle 
    & = \tr_{SE} \left[ e^{it_{n}H}Ae^{-it_{n}H} \Pi_{0} \right] \\
    & = \sum_{k} a_{k} \tr_{SE} \left[ \ket{a_{k}}\!\bra{a_{k}} e^{t_{n}\mathcal{L}} (\Pi_{0}) \right] \\
    & = \sum_{k} a_{k} \mathbb{P}(a_{k},t_{n}) ,
\end{aligned}
\end{equation}
where we have introduced $\mathbb{P}(a_{k},t_{n})$ as the probability of measuring the value $a_{k}$ of the observable $A$ at time $t_{n}$.
An operator expectation value can be seen as a process in which the system evolves up to a time $t_{n}$,
and where an observable $A$ is then projectively measured.
Suppose now the system has a second observable $B=\sum_{j}b_{j}\ket{b_{j}}\!\bra{b_{j}}$,
we can then consider two-time correlation functions of the form
\begin{equation}
\begin{aligned}
    \left\langle A(t_{n})B(t_{m}) \right\rangle = \sum_{k j} a_{k} b_{j} \mathbb{W}(a_{k},t_{n}; b_{j},t_{m}) ,
\end{aligned}
\end{equation}
where in this case we have introduced the function $\mathbb{W}$ (notice that this is not a probability distribution) by
\begin{equation}
\begin{aligned}
    & \mathbb{W}(a_{k},t_{n}; b_{j},t_{m}) = \\
    & \ \ \ \ \ \ \tr_{SE} \left[ \ket{a_{k}}\!\bra{a_{k}} e^{(t_{n}-t_{m})\mathcal{L}} (\ket{b_{j}}\!\bra{b_{j}} e^{t_{m}\mathcal{L}} (\Pi_{0})) \right] .
\end{aligned}
\end{equation}
Two-time correlation functions of this type do not correspond directly to any realizable sequence of measurements,
and it is not generally possible to establish a direct relation to an underlying unique probability distribution
(In general, you can always write these in terms of a linear combination of probabilities, albeit not from measurements in a single basis~\cite{Sakuldee_2018}).
In spite of this, correlation functions find wide application.
For our purposes the crucial thing is that in both cases the function we want to compute can be expressed as unitary evolution on the joint system-environment space,
punctuated by an operation on the system.

The formalism developed below focus on the case where the implemented operations corresponds to generalized measurements,
this approach has the advantage that one can work strictly with positive operators.
The developed method, however, can equally well be applied to more general correlation functions, such as the one considered above.

\subsection{General measurements and the process tensor}
Generally, an instantaneous measurement on the open system is represented mathematically by a completely-positive superoperator $\mathcal{O}^{S}_{x_{n}}$,
which operates only on the system and not on the environment.
Here the index $x_{n}$ labels the set of measurement outcomes of the specific measurement implemented at time $t_{n}$.
For a measurement implemented instantaneously the full state transforms as
\begin{equation}
    \Pi_{n;x_{n}} \ = \ \mathcal{O}_{x_{n}}^{S}\otimes \mathcal{I}^{E} \left( \Pi_{n} \right) .
\end{equation}
In the remainder of the paper we will for convenience not write the identity map on the environment 
or the superscript $S$ on the system superoperator explicitly.
Performing a measurement is a non-deterministic operation,
this means that the resulting state is generally conditioned on the measurement outcome (indicated by the subscript),
and sub-normalized (we take this to be implied by the presence of the subscript).
Throughout we will keep to the operational language in which the considered operations correspond to a generalized measurement.
However, the results we present can straightforwardly accommodate superoperators representing more abstract transformations
useful in the computation of e.g. nonequilibrium Green functions.

Now, consider the statistics obtained by performing a measurement at every discrete timestep.
The probability of realizing a particular measurement record $x_{n:0} = \left\lbrace x_{n},...,x_{0} \right\rbrace$ then takes the form
\begin{equation}
    \mathbb{P}\left( x_{n:0} \right)
    = \tr \left[ \mathcal{O}_{x_{n}} e^{\delta t \mathcal{L}} \mathcal{O}_{x_{n-1}} \ ... \ e^{\delta t \mathcal{L}} \mathcal{O}_{x_{0}} \left( \Pi_{0} \right) \right].
\end{equation}
Notice that even though we are including a measurement at each timestep,
we are not in the final analysis forced to actually implement a measurement,
since the identity map corresponds to a trivial measurement

Later when we present our generalized transfer-tensor formalism,
we will find it useful to note that the above probability can be rewritten by means of the Choi-Jamiolkowski isomorphism \cite{Pollock2018PRA}.
Suppose that we have available a collection of copies of our system.
For reasons which will be clear shortly we semantically divide the set of copies into input systems $R_{j}$ and output systems $S_{j}$.
Given this collection of ancilla systems we introduce the \textit{process tensor} in its many-body Choi state representation \cite{Pollock2018PRA}
\begin{equation}
\begin{aligned}
    \Upsilon_{n:0} = \ 
    & \tr_{E} \left[ e^{\delta t \mathcal{L}_{n}} \ ... \ e^{\delta t \mathcal{L}_{1}} \right. \\
    & \ \ \ \ \ \ \ \ \ \ \left. \left( \Psi_{S_{n}R_{n-1}} \otimes \ ... \ \Psi_{S_{1}R_{0}} \otimes \Pi_{S_{0}E} \right) \right] .
\end{aligned}
\end{equation}
Here the Liouville operator $\mathcal{L}_{j}$ operates on the output space $S_{j}$ and the environment $E$,
and furthermore $\Pi_{S_{0}E}$ denotes the arbitrary initial state of the environment and output space $S_{0}$.
In the above $\Psi_{SR} \equiv \sum_{a,b = 1}^{d} \ket{aa} \! \bra{bb}$ denotes the unnormalized maximally entangled state of ancillary systems $S$ and $R$,
where $d$ is the Hilbert space dimension.
Although the above constitutes the Choi state representation of the process tensor,
we will for simplicity refer to it as the process tensor throughout.
Furthermore we define the Choi state of the \textit{measurement sequence} by
\begin{equation}
\begin{aligned}
    O_{x_{n:0}} = \
    & \mathcal{O}_{x_{n}} \otimes \mathcal{I} \otimes \mathcal{O}_{x_{n-1}} \ ... \ \otimes \mathcal{I} \otimes \mathcal{O}_{x_{0}} \\
    & \left( \Id_{S_{n}} \otimes \Psi_{R_{n-1}S_{n-1}} \otimes \ ... \ \Psi_{R_{0}S_{0}} \right) ,
\end{aligned}
\end{equation}
where $\mathcal{O}_{x_{j}}$ operates on output space $S_{j}$.
Throughout we will refer to this simply as the measurement sequence.
In terms of the process tensor and the measurement sequence,
we can recast the probability of realizing the measurement record $x_{n:0}$ in the form of a spatio-temporal Born rule \cite{Pollock2018PRA}
\begin{equation} \label{eq:mp_prob}
    \mathbb{P}\left( x_{n:0} \right)
    = \tr \left[ O_{x_{n:0}}^{T} \Upsilon_{n:0} \right] ,
\end{equation}
where the trace is over all input and output system spaces.
Given this expression for the probability,
it is clear that the process tensor constitutes a generalization of the open system state to multi-time measurement scenarios.
Crucially for our purposes here,
the process tensor makes it possible to develop a generalization of the transfer-tensor formalism,
without having to deal with the complications of working with states conditioned on particular measurement outcomes.

\subsection{Transfer-tensor formalism for quantum state evolution}
Lastly we introduce the standard transfer-tensor formalism.
Consider the simplest process in which temporal correlations play a role, namely that of a two-time measurement.
Such a process consists of an initial measurement (we call this the preparation procedure) at time $t_{0}$, followed by a second measurement at a later time $t_{n}$.
In this case, the probability of realizing the process defined by the measurement record $\left\lbrace x_{n},x_{0} \right\rbrace$ takes the form
\begin{equation}
\begin{aligned}
    \mathbb{P}(x_{n},x_{0}) 
    = \tr \left[ \mathcal{O}_{x_{n}} e^{n \delta t \mathcal{L}} \mathcal{O}_{x_{0}}\left( \Pi_{0} \right) \right] .
\end{aligned}
\end{equation}
where the maps are applied consecutively.
To compute this probability, we must generally solve the following simulation problem:
First the initial system-environment state must be subjected to the preparation procedure,
and then the resulting conditional state must be evolved up to timestep $n$ according to the Liouville-von Neumann equation.
The statistics associated with the second measurement at timestep $n$,
is then fully characterized by the conditional open system state 
\begin{equation} \label{eq:tp_full_map}
    \rho_{n;x_{0}} = \tr_{E} \left[ e^{n\delta t \mathcal{L}} \mathcal{O}_{x_{0}}\left( \Pi_{0} \right) \right] .
\end{equation}
Computing the open system state by simulating the evolution of the full system-environment state, generally constitutes a highly inefficient computational task.
For this reason it is desirable to develop a dynamical description of the process exclusively within the open system subspace,
which accounts for the effects of the environment on the system dynamics, but otherwise eliminates all information on the environment.

A first step towards such a reduced description,
can be taken by pointing out that any given system-environment state can be decomposed into an uncorrelated part and a correction term.
We take the uncorrelated part to be given by the tensor product of the open system state and an environment reference state $\tau_{E}$ which we take to be independent of time.
The correction term accounts for any system-environment correlations, and for any discrepancy between the reduced environment state and the environment reference state.
It then follows that we can write the initial system-environment state as $\Pi_{0} = \rho_{0}\otimes \tau_{E} + \chi_{0}$, where $\chi_{0}$ is the correction term.
Throughout we will refer to the uncorrelated first term as the product state projection.
If we substitute this decomposition into the right-hand side of Eq.~\eqref{eq:tp_full_map},
then we find that the open system state at time $t_{n}$ consists of a \textit{homogeneous} and an \textit{inhomogeneous} contribution:
\begin{equation}
    \rho_{n;x_{0}} = \mathcal{E}_{n}\left( \rho_{0;x_{0}} \right) + \mathcal{J}_{n;x_{0}} .
\end{equation}
The inhomogeneous term $\mathcal{J}_{n;x_{0}}$ accounts for the discrepancy between the actual initial system-environment state and the product state projection.
We will discuss the problem of accounting for this term towards the end of this section.
The homogeneous contribution is expressed in terms of the dynamical map defined by
\begin{equation} \label{eq:tt_dynamical_map}
    \mathcal{E}_{n} \left( \bullet \right) = \tr_{E} \left[ e^{n \delta t \mathcal{L}} \left( \bullet \otimes \tau_{E} \right) \right] .
\end{equation}
It relates the initial system state to the system state at a later time,
under the assumption that the initial system-environment state is described by the product state projection.

Among many possible definitions \cite{RivasPRL_nonMarkovianity,Breuer_PRL2009_MarkovianityDegree,Pollock2018PRL,Milz_PRL2019_Divisibility},
the presence of temporal correlations in the homogeneous open system dynamics can be related to a violation of the semi-group property $\mathcal{E}_{n} = \mathcal{E}_{1}^{n}$.
A violation of this property implies that the dynamical maps at different times must be computed individually, which is computationally inefficient.
Progress on this problem, can be made by noticing that the set of dynamical maps has an equivalent representation as a set of transfer tensors \cite{Cerrillo2014PRL}.
The transformation rule from one set to the other is the following recursive relation:
\begin{equation} \label{eq:tp_recursion}
    \mathcal{T}_{n} = \mathcal{E}_{n} - \sum_{j=1}^{n-1} \mathcal{T}_{j} \mathcal{E}_{n-j} .
\end{equation}
The homogeneous open system dynamics can equivalently be expressed either in terms of the dynamical maps or the transfer tensors.
In fact the transfer tensors can be seen as the discretized analogue of the Nakajima-Zwanzig memory kernel \cite{Cerrillo2014PRL,Pollock2018tomographically}.

The feature of the transfer-tensor representation that makes it interesting can be appreciated as follows:
Notice that while $\mathcal{T}_{1} = \mathcal{E}_{1}$,
the two-step transfer tensor represents the deviation between the two-step dynamical map
and the dynamical map composed of consecutive applications of the one-step map $\mathcal{T}_{2} = \mathcal{E}_{2} - \mathcal{E}_{1} \mathcal{E}_{1}$.
Generally the transfer tensor $\mathcal{T}_{n}$ encodes the contribution to the dynamical map $\mathcal{E}_{n}$ from temporal correlations in the dynamics extending for a time $n\delta t$.
For an open system dynamics which has a bounded memory time,
we thus expect that the contribution from the transfer tensor $\mathcal{T}_{n}$ to the open system dynamics should approach zero as $n$ increases.
If this is approximately the case,
then the recursive relation Eq.~\eqref{eq:tp_recursion} can be used to iteratively construct approximate dynamical maps extending over arbitrarily long times,
given only a finite set of transfer tensors.
We postpone a discussion of the approximation error to the next section,
where it can be approached for the multi-time case.
Before proceeding we point out that the transfer-tensor representation,
does not free us from having to construct the set of dynamical maps up to a time comparable to the correlation time of the environment.
Rather they provide an efficient representation of the information contained in the constructed maps,
and allows us to efficiently obtain dynamical maps extending over longer times than initially simulated,
if the open system dynamics is finitely correlated in time.

Lastly we must return to the inhomogeneous term.
The problem of accounting for a non-zero inhomogeneous term has been discussed by Buser et al.~\cite{Buser2018PRA}, who proposed the following strategy:
(i) The dynamical maps are generated from a simulation method of choice, and the transfer tensor are reconstructed.
(ii) The reduced state corresponding to correlated initial conditions is simulated using a numerical method capable of dealing with correlated initial conditions.
(iii) The inhomogeneous term is inferred by looking at the difference between the simulated open system state and the homogeneous contribution.
If it is found that the inhomogeneous term vanishes at sufficiently long times,
then the long-time behaviour can be studied exclusively in terms of the homogeneous contribution.

\section{Multi-time measurements}
In the last section we discussed the standard transfer-tensor method for quantum state evolution,
and described how the process tensor is the natural generalization of the quantum state to multi-time measurement scenarios.
In this section we make use of the process tensor,
to construct a generalized transfer-tensor formalism, which is applicable to any multi-measurement process and not only to quantum state evolution.
In addition we analyse the error growth associated with implementing a memory time cutoff, and provide an upper bound on the error.

\subsection{Multi-time transfer tensors}
\begin{figure}
    \begin{center}
	\includegraphics[width=6.5cm]{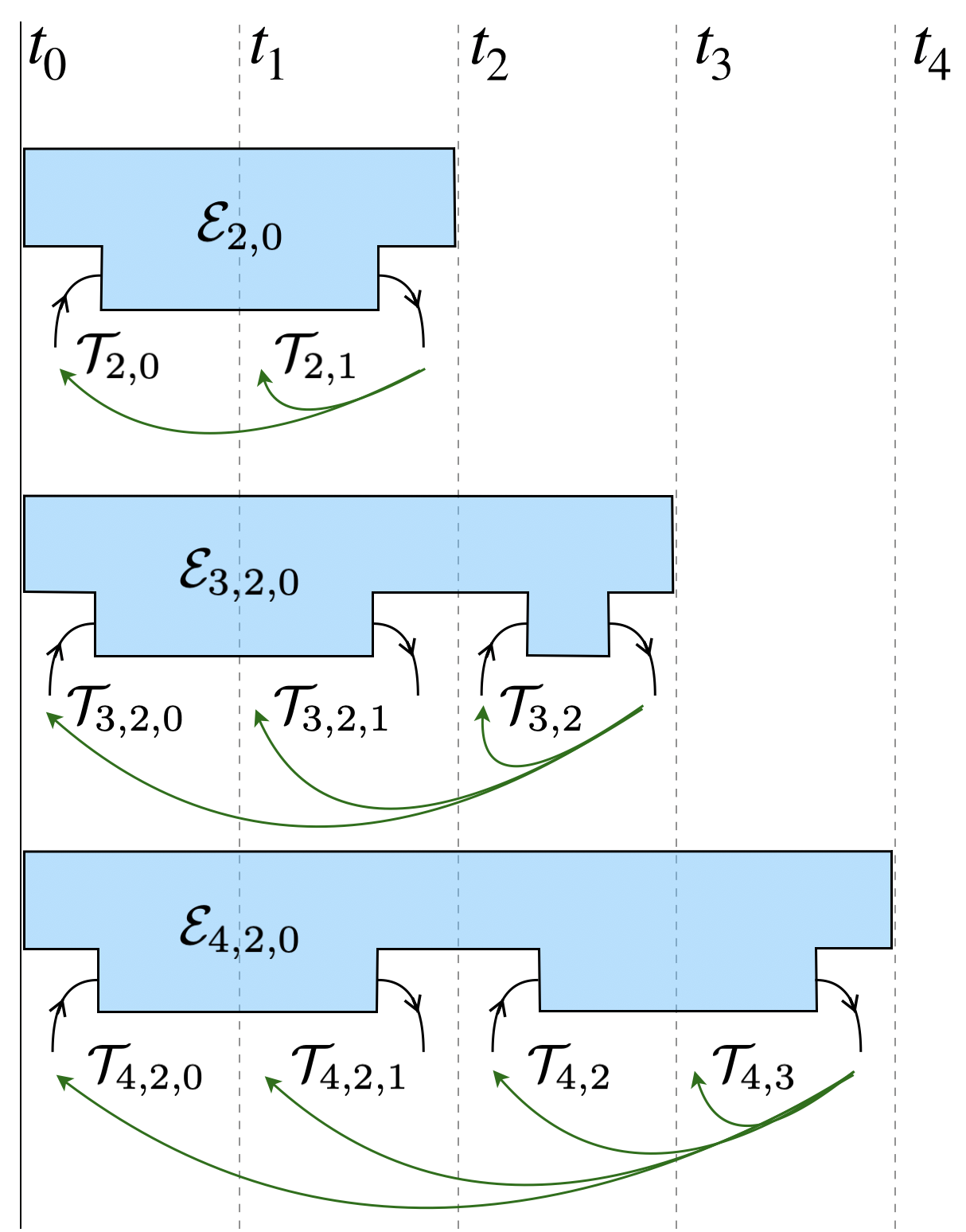}
	\caption{Diagram illustrating the relation between the multi-time dynamical maps and the transfer tensors.
	The transfer tensors decompose the dynamical maps into elements encoding memory effects across a specific timescale.
	The generalized approach includes transfer tensors propagating memory effects across an implemented measurement.
    The holes with input and output arrows signify a timestep at which an operator might act on the system,
    in the case where the implemented operator is an identity (e.g. at time $t_{1}$) this is represented by the absence of a hole.
    Multiple subscripts under the dynamical map symbol indicate the times at which a non-identity manipulation acts on the system.}
    \label{fig:schematic}
    \end{center}
\end{figure}

Similarly to the two-time measurement case,
we write the initial system-environment state as the sum of a product state projection and a correction term.
The process tensor then decomposes into a homogeneous and an inhomogeneous contribution:
\begin{equation} \label{eq:mp_process}
    \Upsilon_{n:0}
    = \mathcal{E}_{n:0} \otimes \rho_{0} + \mathcal{J}_{n:0} .
\end{equation}
The inhomogeneous term $\mathcal{J}_{n:0}$ accounts for any differences between the actual initial system-environment state and the product-state projection.
The homogeneous term is given by the tensor product of the initial system state and the dynamical tensor defined by
\begin{equation} \label{eq:mp_dyn_tensor}
\begin{aligned}
    \mathcal{E}_{n:k} \equiv \
    & \tr_{E} \left[ e^{\delta t \mathcal{L}_{n}}e^{\delta t \mathcal{L}_{n-1}}\ ... \ e^{\delta t \mathcal{L}_{k+1}} \right. \\
    & \ \ \left. \left( \Psi_{S_{n}R_{n-1}} \otimes \ ... \ \Psi_{S_{k+1}R_{k}} \otimes \tau_{E} \right) \right] .
\end{aligned}
\end{equation}
The dynamical tensor provides the multi-time generalization of the dynamical maps introduced for the two-time measurement (see Eq. \eqref{eq:tt_dynamical_map}),
and conceptually we can think of the dynamical tensor as the Choi state representation of a sequence of correlated dynamical maps on the open system space (see Fig.~\ref{fig:schematic}).

We now derive a generalized transfer-tensor representation of the dynamical tensor,
by adapting the approach developed for the two-time transfer-tensor formalism by Pollock et al.~\cite{Pollock2018tomographically}. 
First we define the following projection superoperator acting on the dynamical tensor
\footnote{Notice that due to the causal nature of the dynamical tensor, one can obtain $\mathcal{E}_{j:k}$ from $\mathcal{E}_{n:k}$ by taking the trace over all input and output spaces from timestep $n$ down to the output space at timestep $j$.}
\begin{equation}
    \mathcal{P}_{j} \mathcal{E}_{n:k} \equiv \mathcal{E}_{n:j}\otimes\mathcal{E}_{j:k} \ \ \ \ \text{for} \ k\!<\!j\!<\!n
\end{equation}
and its complement $\mathcal{Q}_{j} \equiv \mathcal{I}-\mathcal{P}_{j}$.
These projection operators are analogous to the standard Nakajima-Zwanzig projectors~\cite{BreuerPetruccione2002},
notice, however, that they are defined directly on the dynamical tensor and not on the underlying $SE$ dynamics
\footnote{In the dilated/underlying picture, the action of the projection operators can be equivalently represented by the usual Nakajima-Zwanzig projectors}.
Making use of the identity, we can iteratively decompose the dynamical tensor as
\begin{equation}
\begin{aligned}
    \mathcal{E}_{n:k} = \ & (\mathcal{P}_{n-1}+\mathcal{Q}_{n-1}(\mathcal{P}_{n-2}+\mathcal{Q}_{n-2}(... )))\mathcal{E}_{n:k} \\
    = \ & \sum_{j=k+1}^{n-1} \mathcal{T}_{n:j} \otimes \mathcal{E}_{j:k} + \mathcal{T}_{n:k}
\end{aligned}
\end{equation}
where we have introduced the set of generalized transfer tensors $\mathcal{T}_{n:k}$.
It is straightforward to show that these are related to the dynamical tensor by the recursive relation
\begin{equation} \label{eq:mp_recursion}
\begin{aligned}
    \mathcal{T}_{n:k}
    = \ & \mathcal{Q}_{n-1}...\mathcal{Q}_{k+1}\mathcal{E}_{n:k} \\
    = \ & \mathcal{E}_{n:k} - \sum_{j=k+1}^{n-1}\mathcal{T}_{n:j} \otimes  \mathcal{E}_{j:k} .
\end{aligned}
\end{equation}
Notice that similarly to the standard case we have for the one-step transfer tensor  $\mathcal{T}_{1:0} = \mathcal{E}_{1:0}$,
while for the two-step transfer tensor $\mathcal{T}_{2:0} = \mathcal{E}_{2:0} - \mathcal{E}_{2:1}\otimes \mathcal{E}_{1:0}$.
Generally the transfer tensor $\mathcal{T}_{n:0}$ quantifies the contribution to the dynamical tensor $\mathcal{E}_{n:0}$
from temporal correlations extending for a time duration $n\delta t$.
In contrast to the standard case, however,
the formulation here allows for experimental interventions at all intermediate times.

\subsection{Long-time propagation and error growth}
Now suppose that a simulation of the dynamical tensor $\mathcal{E}_{n:0}$ can be performed up to a timestep $l$.
Then using the above recursive relation we can iteratively reconstruct the set of transfer tensors $\mathcal{T}_{l:k}$ for all $k\!<\!l$.
This is possible since the open evolution under study is generated by a time-independent Hamiltonian,
which implies that the dynamical tensors $\mathcal{E}_{n:k}$, and consequently the transfer tensors,
become translation invariant in the sense that $\mathcal{E}_{n:k}$ and $\mathcal{E}_{n-k:0}$ are equivalent up to a translation of the system input-output spaces on which they are defined.
For a dynamics finitely correlated in time, we expect that the transfer tensors describing temporal correlations extending for longer than the environment correlation time,
to contribute negligibly to the dynamics of the system.
If we assume that the timestep $l$ is sufficiently large in this sense,
then to a good approximation the homogeneous process can be modelled by the set of truncated dynamical maps
\begin{equation}
    \mathcal{E}_{n:0}^{(l)} = \sum_{k=1}^{l} \mathcal{T}_{n:n-k} \otimes \mathcal{E}_{n-k:0}^{(l)} \ \ \ \ \text{for} \ \ \ \ n\!>\!l .
\end{equation}
The crucial thing here is that given only the initially simulated set of dynamical tensors, and the reconstructed set of transfer tensors,
we can compute approximate dynamical tensors extending over longer times than that initially simulated.

To quantify the error induced by truncating the transfer tensor expansion,
we use the Frobenius norm of the difference between the exact and the reconstructed dynamical tensor.
Writing out the individual terms in the difference, and making use of the triangle inequality and the Cauchy-Schwartz inequality,
we straightforwardly find that the error is upper bounded as
\begin{equation}
\begin{aligned}
		\mid \! \mid \mathcal{E}_{n:0} - \mathcal{E}_{n:0}^{(l)} \mid \! \mid_{F} \ \leq \ \sum_{k=l+1}^{n} (n+1-k) \mid \! \mid \mathcal{T}_{k:0} \mid \! \mid_{F} .
\end{aligned}
\end{equation}
If we take it as a condition for the TTM to be applicable that the transfer tensor norm is monotonically decreasing for times longer than the imposed memory cutoff,
then the error growth is in the worst case quadratic,
with a proportionality constant given by the leading transfer tensor norm $\norm{\mathcal{T}_{l+1:0}}_{F}$.
In many cases this error growth is sufficiently slow to enable accurate predictions of the system steady-state.

\subsection{Generating closed dynamical relations  }
We now show how the standard transfer-tensor formalism can be recovered from the general formulae.
First we consider a measurement sequence consisting of a measurement at time $t_{0}$ and one again at time $t_{n}$.
All other measurements are taken to be identity maps, that is $\mathcal{O}_{x_{j}}= \mathcal{I}$ for all $j$ except $j=0$ and $j=n$.
Then we can take the trace in Eq.~\eqref{eq:mp_prob} over all intermediate output-input spaces,
except for output space $S_{n}$ and the input-output space $R_{0}S_{0}$.
In carrying out the trace we are projecting an output-input space $R_{j}S_{j}$ of the process tensor onto a maximally entangled state $\Psi_{R_{j}S_{j}}$.
From Eq.~\eqref{eq:mp_process} this results in the two-time process tensor
\begin{equation}
    \Upsilon_{n,0} = \mathcal{E}_{n,0} \otimes \rho_{0} + \mathcal{J}_{n,0} ,
\end{equation}
where the inhomogeneous terms is obtained from $\mathcal{J}_{n:0}$ by projections at intermediate times.
The homogeneous contribution is in this case given in terms of the two-time dynamical tensor obtained by subjecting Eq.~\eqref{eq:mp_dyn_tensor} to the projection procedure
\begin{equation}
    \mathcal{E}_{n,k} = \tr_{E} \left[ e^{(n-k) \delta t \mathcal{L}_{n}} \left( \Psi_{S_{n}R_{k}} \otimes \tau_{E} \right) \right] .
\end{equation}
The equivalent two-time transfer-tensor representation is obtained by subjecting the general relation Eq.~\eqref{eq:mp_recursion}
to the same projection onto maximally entangled states of intermediary output-input spaces, that is
\begin{equation}
    \mathcal{T}_{n,m} = \mathcal{E}_{n,m} - \sum_{k=m+1}^{n-1}\mathcal{T}_{n, k} * \mathcal{E}_{ k,m} .
\end{equation}
For convenience, we have introduced the star notation to denote the projection of the intermediate boundary output-input space, that is
\begin{equation}
\begin{aligned}
    \mathcal{T}_{n, k} * \mathcal{E}_{ k,m} \equiv
    \tr_{R_{k}S_{k}} \left[ \Psi_{R_{k}S_{k}} \left( \mathcal{T}_{n,k} \otimes \mathcal{E}_{k,m} \right)\right] .
\end{aligned}
\end{equation}
The above results show, that in considering the two-time measurement, we recover the standard transfer-tensor formalism,
with the trivial difference that instead of relations involving compositions of superoperators,
we have a projection of the intermediate output-input space as a consequence of working within the Choi state representation.

In addition to recovering the standard two-time measurement results,
the generalized formalism makes it possible to study three-time measurement statistics.
We consider a measurement sequence consisting of an initial measurement at time $t_{0}$ followed by a measurement at time $t_{m}$ and one again later at time $t_{n}$.
All other measurements are taken to be identity maps.
Then we take the trace in Eq.~\eqref{eq:mp_prob} over all output-input spaces except for $R_{0}S_{0}$, $R_{m}S_{m}$ and $S_{n}$.
Taking the trace corresponds to projecting intermediate output-input spaces onto maximally entangled states,
and similarly to above this results in the three-time process tensor
\begin{equation}
    \Upsilon_{n,m,0} = \mathcal{E}_{n,m,0} \otimes \rho_{0} + \mathcal{J}_{n,m,0} ,
\end{equation}
where the inhomogeneous terms is obtained from $\mathcal{J}_{n:0}$ by projections at all intermediate times except for time $t_{m}$.
The homogeneous term is expressed in terms of the three-time dynamical tensor obtained by subjecting Eq.~\eqref{eq:mp_dyn_tensor} to the projection procedure
\begin{equation}
\begin{aligned}
    \mathcal{E}_{n,m,k} = \
    & \tr_{E} \left[ \right. e^{(n-m)\delta t \mathcal{L}_{n}} e^{(m-k) \delta t \mathcal{L}_{m}} \\
    & \ \ \ \ \ \ \ \ \left. \left( \Psi_{S_{n}R_{m}} \otimes \Psi_{S_{m}R_{k}} \otimes \tau_{E} \right)\right] .
\end{aligned}
\end{equation}
Furthermore, by subjecting Eq.~\eqref{eq:mp_recursion} to the projection procedure,
it follows that the three-time dynamical tensor has the following transfer-tensor representation
\begin{equation}
\begin{aligned}
    \mathcal{T}_{n,m,k} = \
    & \mathcal{E}_{n,m,k} - \sum_{j=k+1}^{m-1}\mathcal{T}_{n,m,j} * \mathcal{E}_{j,k} \\
    & - \mathcal{T}_{n,m}\otimes \mathcal{E}_{m,k} - \sum_{j=m+1}^{n-1}\mathcal{T}_{n,j} \mathcal{E}_{j ,m,k} .
\end{aligned}
\end{equation}
Notice that we obtain a closed dynamical relation,
in the sense that the three-time dynamical tensors can be decomposed into contributions from the two-time dynamical tensors, the two-time transfer tensors and the three-time transfer tensors.
The relation between the transfer tensors and the dynamical tensor is illustrated in Fig.~\ref{fig:schematic}.

Furthermore, the three-time transfer tensors have the feature that if we want to compute steady-state two-point correlation functions,
we can simply specify a sufficiently large value of $m$ in the above equation.
The equation is closed in the sense that the three-time transfer tensors and the three-time dynamical maps,
need not be computed for any other value of $m$.
This makes it possible to compute steady-state correlation functions by first propagating the system into the steady state,
by a two-time transfer tensors propagation for a sufficiently long-time.
Then an operation is performed on the system (or an arbitrary operator is implemented).
Following the operation the system can be further propagated using the two- and three-time transfer tensors.
Crucially, this can be done while taking full account of the system-environment correlations present in the steady state,
something which has not generally been possible before.
When such calculations are performed, the quantum regression theorem is often applied \cite{BreuerPetruccione2002}, which corresponds to discarding the three-time transfer-tensors.

Lastly we mention the flexibility inherent in the transfer-tensor method in choosing a suitable temporal resolution.
That is, even though we generate dynamical maps over a temporal grid with resolution $\delta t$,
we can define the transfer tensors with respect to a coarser grid, $s \delta t$ for $s$ some positive integer, without increasing the overall simulation error.
This means we can tailor the transfer-tensor decomposition to any desired temporal resolution of the system dynamics.
If the desired resolution has the feature that dynamics is well-captured by the single leading transfer tensor, then we recover a Markovian description of the process.

\begin{figure}
    \begin{center}
	\includegraphics[width=8.5cm]{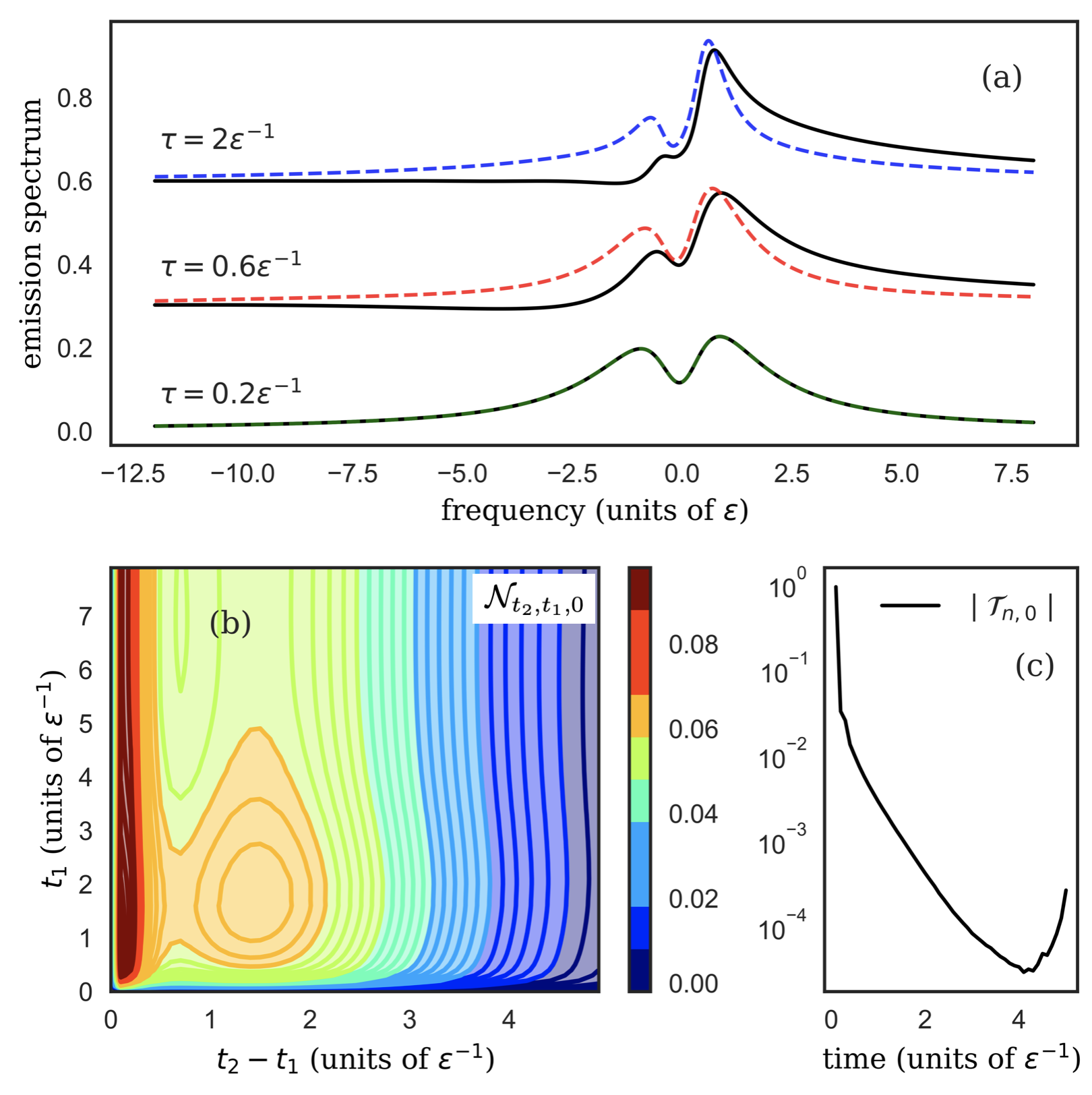}
	\caption{(a) Steady-state phonon emission spectrum for the Ohmic spin-boson model computed using the generalized transfer-tensor method.
	The parameters used are $\alpha = 0.3$, $\delta t = 0.1 \varepsilon^{-1}$, $k_{B}T = 0.1\varepsilon$ and $\omega_{c} = 10\varepsilon$.
	The plot shows the spectra obtained both when including three-time transfer tensors in the propagation (solid lines),
	and the Markovian result obtained when these terms are neglected (dotted lines).
	The time $\tau$ denotes the cutoff time used for the transfer-tensor propagation, for $\tau=0.6\varepsilon^{-1}$ the curve is offset by 0.3 and for $\tau=2\varepsilon^{-1}$ by 0.6.
	(b) Relative entropy measure of non-Markovianity
	(c) Frobenius norm of the two-point transfer tensors. Notice the increase of the transfer tensor norm at long times.
	This effect is well-known in the context of combined TEDOPA-TTM computations \cite{Rosenbach_NJP2016},
	where it signifies a back-reflection of excitations emitted into the environment due to a finite-size effects.
	For the case considered here the transfer-tensor method allows us to remove finite-size effects, by simply neglecting these transfer tensors.}
    \label{fig:spectrum&measure}
    \end{center}
\end{figure}

\section{Application to the spin-boson model}
In this section, we turn to an explicit application of the methods developed above to the paradigmatic spin-boson model~\cite{BreuerPetruccione2002}.
First we apply the generalized transfer-tensor method to study the steady-state emission spectra, and investigate the effects of system-environment correlations present in the steady state.
We then make use of the fact that we are computing the process tensor, to operationally quantify the non-Markovianity of the process~\cite{Pollock2018PRL},
and study the dynamical build-up of system-environment correlations.

\subsection{Steady-state emission spectra}
For concreteness we consider a spin-boson type model with a single two-level system interacting with a bosonic environment.
This system is characterized by the Hamiltonian
\begin{equation}
    H = \frac{\varepsilon}{2} \sigma_{x} + \sigma_{z} \sum_{j} g_{j} \left( a_{j} + a_{j}^{\dagger}  \right) + \sum_{j} \omega_{j} a_{j}^{\dagger} a_{j}
\end{equation}
where $\sigma_{x},\sigma_{y},\sigma_{z}$ are Pauli spin operators, $\varepsilon$ is the energy splitting of the two-level system, $g_{j}$ is a coupling coefficient,
and $a_{j}^{\dagger},a_{j}$ are Bosonic creation and annihilation operators of an environment mode with energy $\omega_{j}$.
The effects of the bosonic environment on the spin dynamics is fully characterized by the spectral density $J(\omega) = \alpha \omega_{c} (\omega/\omega_{c})^{s} e^{-\omega/\omega_{c}}$,
where $\alpha$ is the coupling strength, $\omega_{c}$ is the bath cutoff frequency and $s$ is the Ohmicity.
Two-, three- and multi-time dynamical maps can be efficiently simulated for this model using the TEMPO algorithm and its multi-time generalization \cite{Strathearn2018NC,Jorgensen2019PRL}.
The reference state of the environment is taken to be the thermal state with thermal energy $k_{B}T$,
where $k_{B}$ is Boltzmann's constant.

In many cases we are interested in studying second order correlation functions.
This is the case when for example studying absorption and emission spectra of quantum systems.
In particular we can consider the steady-state emission spectrum given by~\cite{PhysRevA.93.022119}
\begin{equation}
    S(\omega) =  \text{Re} \left[\int_{0}^{\infty} d\tau ( g^{(1)}(\tau)-g^{(1)}(\infty)) e^{-i \omega \tau} \right] ,
\end{equation}
defined in terms of the two-point correlation function 
$g^{(1)}(\tau) = \lim_{t\rightarrow \infty} \left\langle \sigma_{+}(t+\tau)\sigma_{-}(t) \right\rangle$.
The two-point correlation function is defined in terms of the raising and lowering operators on the spin system  $\sigma_\pm = \frac{1}{2}(\sigma_x \pm i \sigma_y)$.
We can compute this function within our three-time measurement scenario,
by implementing the raising and lowering operators in the place of measurement operators.

In Fig.~\ref{fig:spectrum&measure}a we show the steady-state phonon emission spectrum for the spin-boson model computed using the generalized transfer-tensor approach.
The spectrum has been calculated both with (black lines) and without (coloured lines)
including the three-time transfer tensors propagating correlations across the implemented lowering operator.
This corresponds to using an approximate set of transfer tensors, computed via
\begin{equation}
\begin{aligned}
    \mathcal{T}^{\text{reg}}_{n,m,k} = \ & \mathcal{E}_{n,m,k} - \mathcal{T}_{n,m}\otimes \mathcal{E}_{m,k} \\
    & - \sum_{j=m+1}^{n-1}\mathcal{T}_{n,j} \mathcal{E}_{j ,m,k} .
\end{aligned}
\end{equation}
This is equivalent to the quantum regression theorem approach~\cite{BreuerPetruccione2002},
in which it is assumed that no correlations are carried across an implemented operation.
In Fig.~\ref{fig:spectrum&measure}a we show the computed spectra for three different memory cutoff times $\tau$,
at $\tau = 2\varepsilon^{-1}$ the spectrum was converged, such that the black line in this case gives the exact spectrum.

In contrast to previous methods, the multi-time transfer-tensor approach applied here is able to fully account for steady-state system-environment correlations.
This is done without the need to consider inhomogeneous contributions to the dynamics,
as the three-time transfer-tensors makes it possible to homogeneously propagate the state across an implemented operation.
From the figure we observe that these steady-state correlations play an important role in the emission spectrum,
in particular we see that including the three-time transfer tensors tend to suppress the negative frequency part of the spectrum.

\subsection{Quantifying non-Markovianity}
Following the operational approach introduced by Pollock et al.~\cite{Pollock2018PRL},
we define the sub-manifold of process tensor Choi states which can be expressed as product states, as Markovian processes.
This class of processes represent the case where no correlations can be carried across an implemented operation.
Having defined the Markov process, we can define a measure of non-Markovianity $\mathcal{N}$.
We do this using the quantum relative entropy, which for the three-time process tensor gives 
\begin{equation}
\begin{aligned}
\mathcal{N}_{n,m,0}
= \tr \left[ \Upsilon_{n,m,0} \left( \ln \Upsilon_{n,m,0} 
- \ln \Upsilon_{n,m,0}^{\text{markov}} \right) \right] .
\end{aligned}
\end{equation}
Here the Markovian process is given by $\Upsilon_{n,m,0}^{\text{markov}} = \mathcal{E}_{n,m}\otimes \mathcal{E}_{m,0}\otimes \rho_{0}$.
Measuring non-Markovianity as a relative entropy has a clear operational significance as a hypothesis testing procedure.
That is how likely are we to confuse the Markovian process tensor for the non-Markovian process tensor given $\lambda$ repetitions of the process.
The probability of confusing the Markovian and Non-Markovian process tensors is given by $\mathbb{P}_{n,m,0} = \exp \left( -\lambda \mathcal{N}_{n,m,0} \right)$,
such that a larger relative entropy gives a smaller likelihood of confusing the two processes \cite{Pollock2018PRL}.

In Fig.~\ref{fig:spectrum&measure}b we show the relative entropy computed for the ohmic spin-boson model.
Starting out from an initial product state,
we see how system-environment correlations are significant at short times following an intermediate measurement.
At longer times the system-environment correlations present at the time of the intermediate measurement becomes negligible.
This means that if we are interested in processes in which two implemented measurements do not occur to close together in time,
then a Markovian description of the process is adequate.
In the specific case of Fig.~\ref{fig:spectrum&measure}b,
we see that any non-Markovian effects are negligible when the second measurement is implemented at times greater than roughly $4\epsilon^{-1}$ after the firts measurement.

\section{Conclusion}
To conclude, we have presented a generalization of the transfer-tensor formalism to include multi-time measurement scenarios.
The generalization was naturally formulated within the process tensor framework.
We applied our method to the problem of computing steady-state phonon emission spectra for the spin-boson model.
This problem has been addressed before,
the novelty of our computation is that we can take full account of steady-state correlations without the need to compute inhomogeneous terms.
Furthermore we could study the build up of system-environment correlations during the dynamical evolution,
by looking at the relative entropy measure of non-Markovianity.

In the future it would be interesting to investigate whether an operational non-Markovianity measure could be related directly to the transfer-tensor truncation error.
Perhaps, this would make it possible to give notions such a memory strength and process recoverability~\cite{Taranto_arXiv_MemoryStrength}
a direct practical relevance as quantifiers of errors in the simulation of open quantum system dynamics.
In addition, the methods presented here are likely to find many interesting applications.
As an example we mention the study of emission and absorption spectra of organic and inorganic semiconductors,
here one could directly look at the effects of non-Markovianity, coherence and temperature dependence.

\begin{acknowledgments}
The authors would like to thank Jonatan B. Brask for helpful discussions.
MRJ was supported by the Independent Research Fund Denmark.
\end{acknowledgments}

\bibliography{bibliography}
\end{document}